\documentclass[prl,twocolumn,showpacs]{revtex4}
\usepackage[xdvi]{graphicx}%
\usepackage{dcolumn}
\usepackage{amsmath}
\makeatletter

\def\btt#1{\texttt{\@backslashchar#1}}%
\DeclareRobustCommand\bblash{\btt{\@backslashchar}}%
\makeatother

\newcommand{\bv}[1]{{\boldsymbol #1}}
\newcommand{\lgc}{\left <}
\newcommand{\rgc}{\right >}
\newcommand{\fr}[2]{\frac{#1}{#2}}

\begin{document}
\title{Finite-size scaling for non-linear rheology of fluids confined in a small space}
\author{Michio Otsuki }
\email{otsuki@jiro.c.u-tokyo.ac.jp}

\affiliation
{Department of Pure and Applied Sciences,  
University of Tokyo, Komaba, Tokyo 153-8902, Japan}

\date{\today}

\begin{abstract}
We perform molecular dynamics simulations 
in order to examine the rheological transition 
of fluids confined in a small space.
By performing finite-size scaling analysis, 
we demonstrate that this rheological 
transition results from the competition between 
the system size and the length scale of cooperative particle motion.
\end{abstract}


\maketitle


%
%

When fluids are confined in a small space,
their rheological property changes drastically from 
that of the fluids in the bulk.
A typical example of this rheological transition is 
observed in a fluid confined between solid walls. 
When the distance
between the walls is so large that the fluid is considered in the bulk,
it behaves as a normal Newtonian fluid. On the other hand, when the 
distance becomes equivalent to that of approximately eight molecular layers, 
the fluid exhibits an increase 
in its viscosity, the shear-thinning behavior and the appearance of the yield 
stress as well as the enhancement of the relaxation time 
\cite{granick:review}.
This rheological transition
has been extensively  studied by 
experiments \cite{granick:experiment,yamada} and simulations 
\cite{thompson, jabbarzadeh, Klein, Cui1, Cui2}.

%
%

Some researchers have conjectured that this rheological transition
results from structural transition \cite{jabbarzadeh, Klein, Cui1}.
However, in some simulations of fluids confined between walls,
the rheological transition is observed without any clear structural 
transition \cite{thompson}.  Hence, we consider that there is another mechanism
responsible for this rheological transition.

%
%

Let us recall that
glassy materials, such as dense colloidal suspensions and 
super-cooled liquids, also display rheological transition
similar to that observed in fluids confined in a small space
when the temperature decreases or the density 
increases \cite{colloid,supercooled,berthier}.
From this fact, we conjecture that
fluids confined in a small space and glassy materials have common features 
\cite{granick:experiment,yamada}.

%
%

In particular, in glassy materials, it has been observed that the length 
scale of cooperative particle motion, which is called the dynamical 
correlation 
length, becomes comparable with the system size near glass transitions
\cite{glotzer,berthier:length}. Therefore, in the case of fluids confined
in a small space,
the dynamical correlation length 
might be comparable with the system size when the size is sufficiently small.
Our conjecture suggests that the competition between
the dynamical correlation length and the system size is an
origin of the rheological transition.

In order to test whether the rheological transition is actually caused by
the conjectured effect, we investigate a model 
that does not exhibit any structural transition.
By performing numerical simulations with 
a finite-size scaling analysis \cite{goldenfeld}, we demonstrate that 
the rheological transition occurs because of the 
competition between the system size and the dynamical correlation length.

\paragraph{Model:}

%
%

The system is a 80:20 mixture of Lennard-Jones
particles of types A and B in a two-dimensional square box, with an interaction
potential
\begin{equation}
V_{ {\rm \alpha \beta}}({\bv r}) = 4 \epsilon_{{\rm \alpha \beta}} 
\left [ 
\left (
\frac{\sigma_{{\rm \alpha \beta}}}{r}\right)^{12} - 
\left (
\frac{\sigma_{{\rm \alpha \beta}}}{r}\right)^{6} 
 \right ],
\end{equation}
where ${\rm \alpha}$ and ${\rm \beta}$ refer to the two species A and B.
In order to restrict crystallization, we investigate the binary mixture.
The particles have equal masses, and the interaction parameters are 
$\epsilon_{{\rm AB}}=1.5 \epsilon_{{\rm AB}}$, $\epsilon_{{\rm BB}}=0.5
\epsilon_{{\rm AA}}$, 
$\sigma_{{\rm AB}}=0.88 \sigma_{{\rm AA}}$, and  $\sigma_{{\rm BB}}=0.8
\sigma_{{\rm AA}}$.  
The length, energy, and time units are expressed in the standard
Lennard-Jones units: $\sigma_{{\rm A A}}$ (particle diameter),
$\epsilon_{{\rm AA}}$ (interaction energy), and $\tau_0=(m_{{\rm A}}\sigma_{\rm AA}^2
/\epsilon_{{\rm AA}})^{1/2}$, where $m_{{\rm A}}$ is the particle mass. 
The subscript A refers to the major species. 
We fix the total number density $\rho=N/L^2=1.13$, where L is the system size
and $N$ is the number of the particles. 

%
%

The evolution equations of the $i$-th particle's position 
${\bv r}_i=(x_i,y_i)$ and
velocity ${\bv v}_i=(u_i, v_i)$ are numerically integrated by using a leapfrog 
algorithm 
with a time step $\delta t = 0.005$. Here, in order to maintain the temperature 
$T$, we use the Nose-Hoover algorithm \cite{evans}.
In order to realize the shear flow $(\gamma y, 0)$
without structural ordering near the boundaries,
Lees-Edwards boundary conditions are imposed \cite{evans}.
Here, $\gamma$ is the shear rate.

\paragraph{Equilibrium Properties:}

%
%

\begin{figure}[htbp]
\begin{center}
\includegraphics[height=15em]{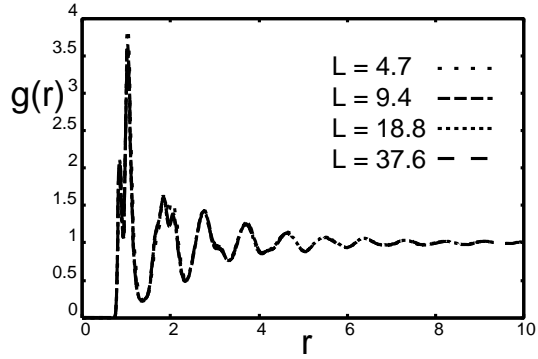}
\caption{
Pair distribution function $g(\bv{r})$ for several values of $L$ when $T=0.8$
and $\gamma=0$.
}
\label{gr}
\end{center}
\end{figure}

We first calculate the pair distribution function defined as
\begin{eqnarray}
g(\bv{r}) = \frac{L^2}{N^2} \sum_{i\neq j} \left < \delta(\bv{r} - 
\bv{r_i}+\bv{r_j}) \right >_{T,\gamma,L}.
\end{eqnarray}
Here, $ \lgc \cdot \rgc_{T,\gamma,L}$ represents the ensemble average 
under the condition that
temperature $T$, the shear rate $\gamma$, and the system size $L$ are provided.
In Fig. \ref{gr}, we show the equilibrium pair distribution functions
for several values of $L$ when $T=0.8$ and $\gamma=0$. 
It can be seen that the structural transition does not occur 
when the system size is decreased.

%
%

\begin{figure}[htbp]
\begin{center}
\includegraphics[height=15em]{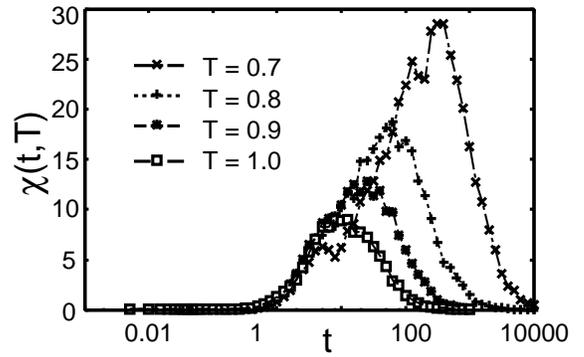}
\caption{
$\chi(t,T)$ as a function of $t$ for several values of $T$ when $L = 37.6$.
}
\label{chi}
\end{center}
\end{figure}

Next, we study the dynamical correlation length $\xi(T)$, 
following Ref. \cite{glotzer,berthier:length}.
For this purpose, we define the local indicator of configuration change as
\begin{eqnarray}
F(\bv{r},t) = \sum_i \delta (\bv{r} - \bv{r}_i(0)) W( \bv{r}_i(t)
- \bv{r}_i(0) ),
\end{eqnarray}
with
\begin{eqnarray}
W(\bv{r}) = 
\left\{ \begin{array}{ll}
1 & (|\bv{r}| < a), \\
0 &  ({\rm otherwise}).\\
\end{array} \right.
\end{eqnarray}
Here, we choose the value of $a$ as $0.3$, which is slightly larger
than the square root of the plateau value of the mean square displacement,
as discussed in Ref. \cite{glotzer}.
Note that $\left < F(\bv{r},t) \right >$ is akin to the intermediate scattering function
when the system is homogeneous.
The correlation function $C(\bv{r},t,T)$ of $F(\bv{r},t)$ is
defined as 
\begin{equation}
C(\bv{r},t,T)  =  \fr{V^2}{N^2}\lgc F(\bv{r},t) F(\bv{0},t) \rgc_{T,\gamma,L}
- \fr{V^2}{N^2} \lgc F(\bv{r},t)\rgc^2_{T,\gamma,L}.
\end{equation}
From $C(\bv{r},t,T)$, the amplitude of the 
correlation is defined as
\begin{eqnarray}
\chi(t,T) & = & \fr{1}{k_B T} \int d\bv{r} C(\bv{r},t,T).
\end{eqnarray}
In Fig. \ref{chi}, we show the time dependence of $\chi(t,T)$ for several
values of $T$ when $L = 37.6$ and $\gamma=0$. $\chi(t,T)$ has a 
maximum value at a time $t_{{\rm max}}$.
As the temperature decreases, $t_{{\rm max}}$ and 
the maximum value $\chi(t_{{\rm max}},T)$ increases.

%
%

\begin{figure}[htbp]
\begin{center}
\includegraphics[height=15em]{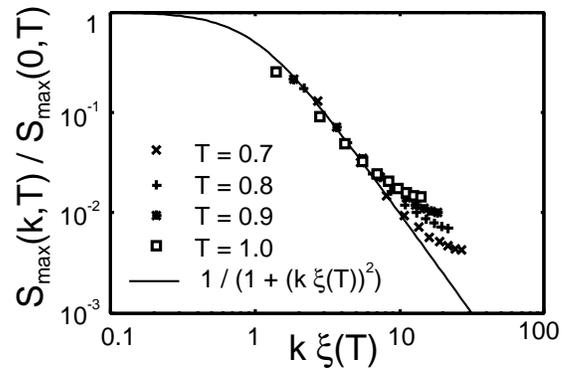}
\caption{
$S_{{\rm max}}(k,T)/S_{{\rm max}}(0,T)$ as a function of $k \xi(T)$ 
for several values of $T$ when
$L = 37.6$.
}
\label{s4maxred}
\end{center}
\end{figure}

\begin{figure}[htbp]
\begin{center}
\includegraphics[height=15em]{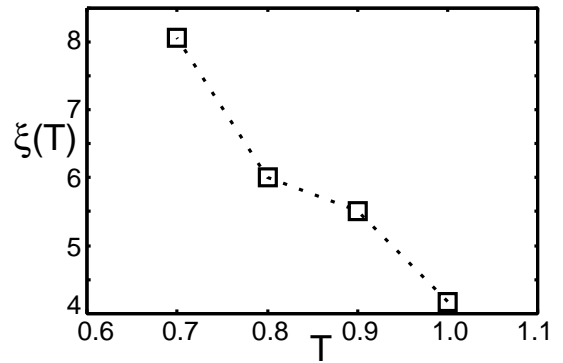}
\caption{
$\xi(T)$ as a function of $T$ for $L = 37.6$.
}
\label{xi}
\end{center}
\end{figure}

In order to extract $\xi(T)$,
we define the Fourier transformation $S(\bv{k},t,T)$ of 
$C(\bv{r},t,T)$ as
\begin{eqnarray}
S(\bv{k},t,T) & = & \int d \bv{r} C(\bv{r},t,T) \exp 
(-i \bv{k} \cdot \bv{r}),
\end{eqnarray}
where
\begin{eqnarray}
S(0,t,T) & = & k_B T \chi(t,T).
\end{eqnarray}
From $S(t,k,T)$, we define $S_{{\rm max}}(k,T) = S(k,t_{{\rm max}},T)$. 
By assuming the functional form of $S_{{\rm max}}(k,T)$ as
\begin{eqnarray}
S_{{\rm max}}(k,T) = \fr{S_{{\rm max}}(0,T)}{1+(k \xi(T))^2},
\label{reduce}
\end{eqnarray}
we estimate $\xi(T)$ for the case that
$\gamma=0$ and $L=37.6$.
In Fig. \ref{s4maxred},
$S_{{\rm max}}(k,T)/S_{{\rm max}}(0,T)$ 
is displayed as a function of $k \xi(T)$. In Fig. \ref{xi}, we show
$\xi(T)$ as a function of $T$.

\paragraph{Rheological Properties:}

%
%

\begin{figure}[htbp]
\begin{center}
\includegraphics[height=15em]{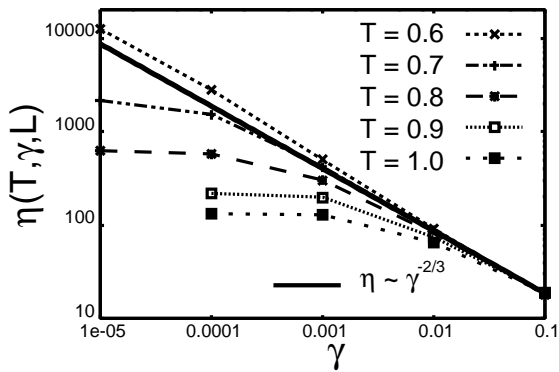}
\caption{
Relation between $\eta$ and $\gamma$ for several values of $T$ when $L=37.6$.
}
\label{eta}
\end{center}
\end{figure}

As a typical rheological property, we investigate the viscosity.
In our system, the shear stress $\sigma_{xy}(T,\gamma,L)$ 
is expressed as
\begin{equation}
\sigma_{xy}(T,\gamma,L) = \frac{1}{L^3} \left< - \sum_{i=1}^N m_i u_{i} v_{i}
+  \sum_{i \neq j} \frac{x_{ij}y_{ij}}{2r_{ij}}
\frac{\partial V(r_{ij})}{\partial r_{ij}} \right >_{T,\gamma,L},
\end{equation}
where $x_{ij} = x_i-x_j$, $y_{ij} = y_i-y_j$, and $r_{ij} = |\bv{r}_i-\bv{r}_j|$
\cite{evans}.
From $\sigma_{xy}(T,\gamma,L)$ and $\gamma$,
the viscosity $\eta(T,\gamma,L)$ is defined as
\begin{eqnarray}
\eta(T,\gamma,L) = \frac{\sigma_{xy}(T,\gamma,L)}{\gamma}.
\end{eqnarray}
In Fig. \ref{eta}, we show 
$\eta(T,\gamma,L)$ as a 
function of $\gamma$ for several values of $T$ when $L=37.6$.
For $T=1.0$, the viscosity is independent of the shear rate
when $\gamma < 0.001$. This regime is called a Newtonian regime.
As the shear rate increases,
the viscosity decreases as $\eta \sim \gamma^{-\alpha}$. 
When the temperature decreases, the Newtonian regime becomes narrower and 
finally disappears.
The exponent $\alpha=2/3$ is observed in experiments
\cite{granick:experiment} and simulations \cite{thompson} 
of thin liquid films.
Furthermore, this exponent is observed in 
simulations of glassy materials \cite{berthier}.

%
%

\begin{figure}[htbp]
\begin{center}
\includegraphics[height=15em]{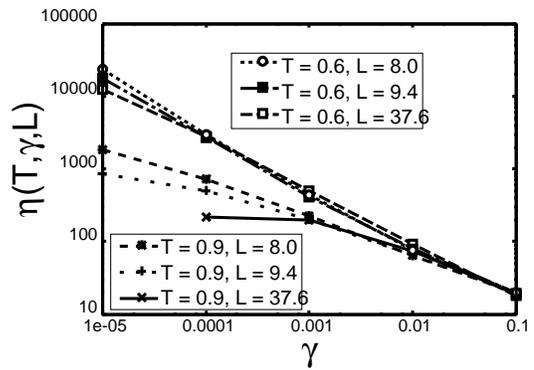}
\caption{
Relation between the viscosity $\eta(T,\gamma,L)$ and the shear 
rate $\gamma$ for several values of $L$ when $T=0.6$ and $0.9$.
}
\label{figure6}
\end{center}
\end{figure}

Now, we investigate the system size dependence of the rheological property.
As examples, in Fig. \ref{figure6}, we show $\eta(T, \gamma, L)$
as a function
of $\gamma$ for several values of $L$ when $T=0.6$ and $0.9$.
At $T=0.6$, $\eta$ changes slightly.
On the other hand, 
when $T=0.9$, $\eta$ strongly depends on $L$ in the low shear rate regime
as in experiments \cite{granick:experiment,yamada} and simulations 
\cite{thompson, jabbarzadeh, Klein, Cui1, Cui2}
of a thin liquid film.

\paragraph{Finite-size scaling:}

%
%

We perform a finite-size scaling analysis in a similar manner to that used for
critical phenomena \cite{goldenfeld}. We first assume the scaling form of the 
viscosity as
\begin{eqnarray}
\eta(T,\gamma,L) = \eta_0(T,L) G\left(\gamma \cdot 
\eta_0(T,L)^{\frac{3}{2}} \right), \label{scale1}
\end{eqnarray}
where $\eta_0(T,L)$ is the Newtonian viscosity
defined as $\eta_0(T,L)=\lim_{\gamma \rightarrow 0} \eta(T,\gamma,L)$.
$G(x)$ is a function having the properties
\begin{eqnarray}
\lim_{x \rightarrow 0} G(x) &=& 1,\\
\lim_{x \rightarrow \infty} G(x) &\sim& x^{-\frac{2}{3}}.
\label{2/3}\\
\end{eqnarray}
It must be noted that this scaling form can be derived from a theory for
the rheological property in glassy materials \cite{otsuki}.
Here, we only consider the case where $\eta_0(T,L)$
has a finite value. 
Next, we assume the functional form of $\eta_0(T,L)$ by $\xi(T)$ and $L$ as
\begin{eqnarray}
\eta_0(T,L) = \eta_I (T) H(\xi(T)/L), \label{scale2}
\end{eqnarray}
where $\eta_I(T)$ is the viscosity in the thermodynamical limit 
$L \rightarrow \infty$ and $H(x)$ is the function having the property
\begin{eqnarray}
\lim_{x \rightarrow 0} H(x) &=& 1.
\end{eqnarray}

%
%

\begin{figure}[htbp]
\begin{center}
\includegraphics[height=15em]{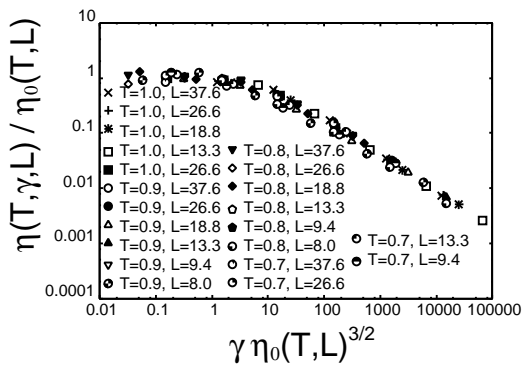}
\caption{
$\eta(T,\gamma,L)/\eta_0(T,L)$ as a function of
$\gamma \cdot \eta_0(T,L)^{3/2}$ for various values of $T$ and system size $L$.
}
\label{etar}
\end{center}
\end{figure}

\begin{figure}[htbp]
\begin{center}
\includegraphics[height=15em]{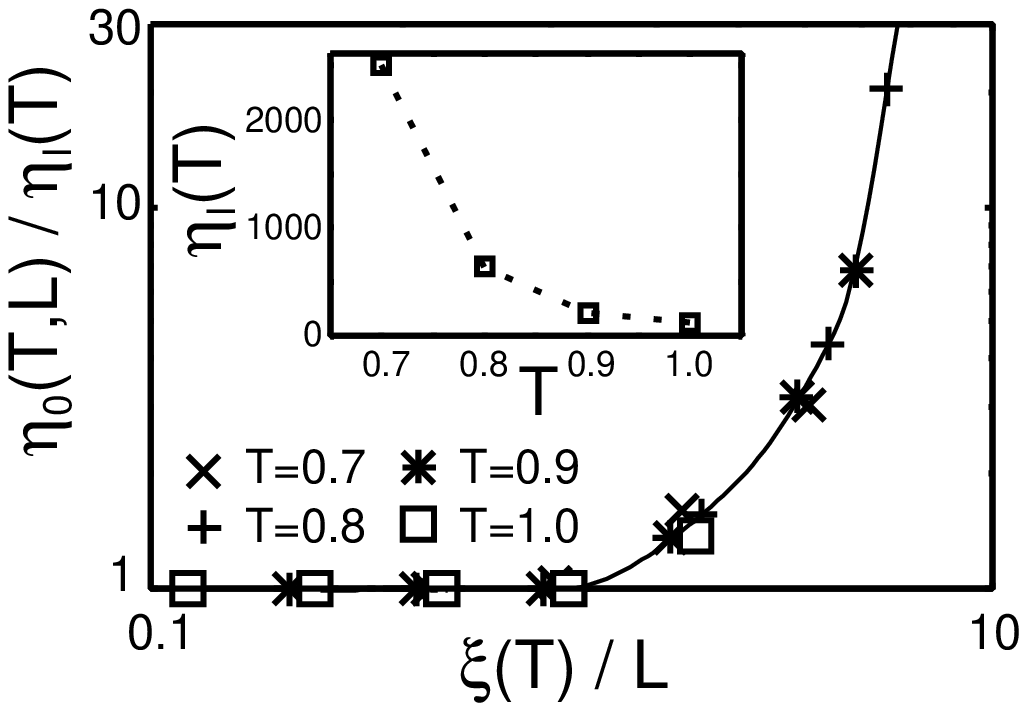}
\caption{
Main : $\eta_0(T,L)/\eta_I(T)$ as a function of $\xi(T) / L$.
Inset : $\eta_I (T)$ as a function of $T$.
}
\label{figure8}
\end{center}
\end{figure}

Then, we test the scaling assumptions described by Eqs.
(\ref{scale1}) and (\ref{scale2}). We show $\eta(T,\gamma,L)/\eta_0(T,L)$
as a function of $\gamma \cdot \eta_0(T,L)^{3/2}$ in Fig. \ref{etar}.
In addition, we show
$\eta_0(T,L)/\eta_I(T)$ as a function of $\xi(T) / L$ in Fig. \ref{figure8}.
These figures clearly indicate that the scaling assumptions described by Eqs. 
(\ref{scale1}) and (\ref{scale2}) are plausible. In addition, the viscosity
$\eta_I(T)$ in the thermodynamical limit is shown in the inset of Figs. 
\ref{figure8}.
From these results, we conclude that the size dependence of the rheological
property in our system results from the competition between 
$L$ and $\xi(T)$.

\paragraph{Conclusion and Discussion:}

%
%

We performed the molecular dynamics simulations of the fluid confined in a small
space with the Lees-Edwards boundary conditions. In these simulations, we 
observed the strong size dependence of the rheological property. In addition, we
numerically calculated the correlation length $\xi(T)$ of the local 
configuration change. 
From the numerical confirmation of the scaling relations given by
Eqs. (\ref{scale1}) and (\ref{scale2}), it has
been concluded that the strong size dependence of the rheological property is 
caused by the competition between the system size and the correlation length.

%
%

Although we have shown that the proposed mechanism affects the rheological 
property in our system without wall boundaries, 
it is not clear whether it would be really effective in experiments of a 
thin fluid film with wall boundaries. In such a situation,
other effects such as the structural transition might be important for
determining the rheological properties. 

%
%

In Ref. \cite{kim},
it is shown that the density relaxation time
of glassy materials increases when the system size becomes comparable with a 
dynamical correlation length. With the assumption that the relaxation time
is proportional to the viscosity \cite{yamamoto}, 
the result in Ref. \cite{kim} is similar to the result described by Eq. 
(\ref{scale2}) and Fig. \ref{figure8}.
However, we wish to note that in Ref. \cite{kim}, the rheological property was 
not studied, and the finite size scaling was not performed.

%
%

Finally, let us remind that the correlation length of the local indicator of 
the local configuration change, which is the dynamical quantity,
 is used in our finite-size scaling analysis.
This choice is natural because the viscosity is approximately calculated from 
the time correlation function of the local configuration change \cite{miyazaki}.
On the other hand, when we recall the fact that the viscosity is 
calculated from the shear stress, which is the static quantity,
it is also expected that the correlation of a static quantity becomes involved
in  the 
scaling analysis.  However, as far as we have studied, we could not find such a 
quantity. This problem will be studied in future.


We thank S. Sasa and K. Hukushima for their valuable advice, and H. Tasaki,
H. Hayakawa, and T. Hatano for their useful comments.

\end{document}